\documentclass[12pt]{article}
\usepackage{lingmacros}
\usepackage[english]{babel}
\usepackage{graphicx}

\large \baselineskip = 20 true pt

\begin{document}
\begin{center}
{\Large \bf Using the decision support algorithms combining different security policies} \vspace{0.5cm}
\end{center}

\begin{center}
S.V. Belim, N.F. Bogachenko, Y.S. Rakitskiy, A.N. Kabanov\\
Dostoevsky Omsk State University, Omsk, Russia

 \vspace{0.5cm}
\end{center}

\begin{center}
{\bf Abstract}
\end{center}

During the development of the security subsystem of modern information systems, a problem of
the joint implementation of several access control models arises quite often. Traditionally,
a request for the user's access to resources is granted in case of simultaneous access permission
by all active security policies. When there is a conflict between the decisions of the security
policies, the issue of granting access remains open. The proposed method of combining multiple
security policies is based on the decision support algorithms and provides a response to the
access request, even in case of various decisions of active security policies. To construct
combining algorithm we determine a number of weight coefficients, use a weighted sum of the
clearance levels of individual security policies and apply the analytic hierarchy process. The
weight coefficients are adjustable parameters of the algorithm and allow administrator to manage
the impact of the various security rules flexibly.

\section{Introduction}

The problem of combining different security policies arises quite often in the administration
of computer systems. For example, consider a database management system based on Windows family
of operating systems. A role-based security policy is the most common in the database management
systems, but the data is stored in files, access to which is controlled by the operating system.
A discretionary security policy is basic for the operating systems, but at a certain level,
a mandatory security policy is realized. Thus, conjugation of three different security policies
is required.

If we consider the different standards of information security in automated systems, we could
see that they also imply the existence of more than one security policy.

The possibility of combining discretionary and mandatory security policies is provided
in the vast amount of information security standards. According to the Orange Book, a computer
system using only discretionary access control belongs to one of the division C classes,
while the addition of mandatory security policy allows it to qualify for a higher class
of division B. Moreover, the Orange Book implies precisely the addition of mandatory
security policy with retention of discretionary security policy capabilities.

The standard approach is to find a perfect solution so that the settings of one security policy
do not conflict with other security policies. To date, there exists a number of different
solution.

In \cite{b1} the expansion of discretionary Take-Grant model, taking into account the mechanism
of mandatory access control, is considered. The article \cite{b2} proposed a universal language
that allows to describe and implement a global integrated security policy for the system,
consisting of a variety of IT environments, each has its own security model and management
domain. This language is implemented by the event monitor. Resolving conflicts between
the security policies is realized by an explicit call of the administrator for the adoption
of the priority solution. In \cite{b3} the authors proposed to extend the security matrix
of discretionary security policy to the security cube that in addition to traditional subjects
and objects has a third dimension - users or groups of users. This additional dimension allows
organizing a mechanism of group access control while remaining within the discretionary security
policy. The paper \cite{b4} is devoted to the joint implementation of the role-based and the
mandatory access control concepts. Theoretical-graph approach to the security lattice
of the mandatory security policy allows combining the requirements of the role-based
and the mandatory models for digraph of entities of computer system. The principal possibility
of creating a security policy, which includes a mandatory and a role-based access control,
is shown. The paper \cite{b6} presents a model of access control for working with XML-documents.
This model combines the advantages of the role-based and the mandatory access control policies.
In particular, it is proposed to use not access control lists but an approach based on security
labels to determine the access rights.

However, there is a fundamental problem of the ideal approach realization consisting
in the absence of evidence that there is a perfect solution. Moreover, the practical
realization of the simultaneous administration of multiple security policies shows that
it is not always possible to make adjustments to ensure the proper functioning of the system.

This article deals with a new approach in combining multiple security policies in a single
computer system based on decision algorithms.

\section{Statement of the problem and a common approach to the solution}

Let us consider a system in which both discretionary and mandatory security policies are
implemented. According to conventional wisdom, which was embodied in virtually all information
security standards, the mandatory security policy provides a high level of information protection
and dominate the discretionary security policy that provides a basic level of data protection.
When there is a conflict between the two settings of security policies, two approaches are used
traditionally. According to the first approach, access is denied if it is prohibited at least
by one security policy. In the second case, the mandatory security policy dominates, and the
decision about access permission is received based on its settings. The first approach can easily
lead to the complete failure of the system, the second approach virtually shut down discretionary
security policy. Furthermore, the mandatory security policy is focused on the overall system. The
administrator defines the security labels of system subjects and objects that can be changed only
by changing the state of the system as a whole but not in a particular access. However,
exceptions are possible. For example, the administrator need to allow access of a specific
subject to a specific object, contrary to the mandatory security policy. The administrator
monitors the contents of the object and can guarantee no leakage of information through
the subject, but cannot guarantee no leakage through other subjects with the same level
of access. This permission may be implemented with the introduction of some additional
security labels for each case, which leads to a significant increase and entanglement
of the security lattice and, consequently, to the complexity of system administration.
Another approach is to use discretionary security policy, which in one case should dominate
the mandatory security policy. In other words, the superstructure can be implemented in the
system, and it will decide on the dominance of a particular security policy in each case.

We formulate the statement of the problem more strictly. Let the two security policies operate
in the system and take decisions based on P1 and P2 algorithms. It is necessary to implement
the algorithm P for taking a decision about what kind of security policy will be used for each
request for access. It should be noted that using the algorithm $P$ is really needed only when
the decisions of $P_1$ and $P_2$ are contradictory.

Traditionally, within the security policy, a decision taken on the access request is the value
from the set $\{0, 1\}$, where zero corresponds to an access denial, and one corresponds to an
access permission. To make a decision about accessibility we extend the range of the algorithm
values to the set $\{-m,..., -1, 0, 1, ..., m\}$ where $m$ is the positive integer.
The values of this range is called the clearance level, it is denoted by the letter $p$.
Access is granted if $p > 0$. The clearance level refers to the probability of information
leakage for a given access:
\[
P(p)=0.5-\frac{p}{2m}.
\]
The higher is the clearance level, the higher is the level of confidence in access.
With this approach, algorithm $P$ takes a decision based on the clearance levels
of individual security policies $p_1$ and $p_2$.

Consider a system in which security policy $P_1$ dominates security policy $P_2$.
We introduce the dominance coefficient $r$, showing how many times the decision taken by
$P_1$ is more important than the decision taken by $P_2$. In this case, the final decision
can be calculated as a weighted sum of the decisions of the two security policies:
\[
p=\frac{r}{r+1}p_1+\frac{1}{r+1}p_2.
\]
Equivalence of the security policies is achieved with $r = 1$. Access is granted if $p > 0$.
Note that $p$ is not necessarily an integer: $p\in [-m, m]$.

\section{The combination of the mandatory and discretionary security policies}

We consider the most common case of combining the mandatory and discretionary security policies.
For the mandatory security policy, we restrict ourselves to the simplest case of a linear
security lattice with $l$ levels of security. In this case, the clearance level can be found
as the difference between the level of confidence of the subject $C(S)$ and the level
of secrecy of the object $C(O)$:
\[
p_1=(C(S)-C(O))\frac{m}{l}.
\]
For the discretionary security policy, the clearance level can be arbitrarily set by the
administrator for each access. If the administrator wants to set a top priority for the access,
he appoints $p_2 = m$. Therefore, only the case of appointment of the default clearance level
is consider. We assume that the total number of possible types of access equal to $M$.
Let the subject requests access to the object for several types of access. In case
of access denial, we assume that
\[
p_2=-k\frac{m}{M},
\]
where $k$ is the number of denied accesses from the list of requested accesses.
If access is granted, then let
\[
p_2=h\frac{m}{M},
\]
where $h$ is the number of granted but not requested types of access.

{\bf Example 1.} Consider the model example of such a system. Suppose that the mandatory security
policy is based on a linear security lattice with four levels $L = \{0, 1, 2, 3\}$.
For discretionary security policy, we assume that four types of access are defined:
$R = \{r, w, a, f\}$. Suppose that at some point in time a request for access $(s, o, r)$
is received, which has the corresponding cell $M[s, o] = \{r, w, a\}$ in the access matrix,
the level of secrecy of the object is $C(o) = 2$, the level of confidence of the subject is
$C(s) = 1$. In this case, $p_1 = C(s) - C(o) = -1$, $p_2 = 2$. When security policies are equal,
then $p = (p_1 + p_2) / 2 = 1.5 > 0$, i.e. access is granted despite the prohibition of the
mandatory security policy. If we increase the priority of the mandatory security policy 3 times,
then:
\[
p=\frac{3}{4}\cdot(-1)+\frac{1}{4}\cdot2=-\frac{1}{4}<0.
\]
In this case, access is denied.

\section{Nonlinear security lattice}

When combining the mandatory and the discretionary security policies, we considered the
simplest case of a linear security lattice with L levels of security. However, in real systems,
the mandatory security policy can be set by nonlinear security lattice, i.e. a variety of
security labels will be partially ordered. With this approach to the implementation of security
policies, the level of confidence of the subject $C(S)$ and the level of secrecy of the object
$C(O)$ may be incomparable. In this case, it is impossible to define the clearance level as the
difference between the level of confidence of the subject $C(S)$ and the level of secrecy of the
object $C(O)$. This means that we need a different approach to determining the clearance level
given by the mandatory security policy.

The classical model of mandatory security policy defines an operator that specifies for each pair
of elements $l_1$ and $l_2$ from the basic set of security levels L a single element of the least
upper bound:
\[
sup(l_1, l_2) = l \Leftrightarrow  (l_1,l_2\leq l)\wedge (\forall l'\in L: (l'\leq l) \Rightarrow (l'\leq l_1 \vee  l'\leq l_2)).
\]
We introduce the operator $dif(l_1, sup(l_1, l_2))$ showing the "distance" from the security
level $l_1$ to the least upper bound of the security levels $l_1$, $l_2$:
\[
dif(l_1, sup(l_1, l_2)) = sup(l_1, l_2) - l_1.
\]
This approach is possible because the lattice elements $l_1$ and $sup(l_1, l_2)$ are by
definition always comparable. This operator allows determining the number of levels of the
security lattice from element $l_1$ to $sup(l_1, l_2)$. Note that this value is always
non-negative.

We determine the clearance level p1 for incomparable in the lattice level of confidence of the
subject $C(S)$ and the level of secrecy of the object $C(O)$ as a negative modulus of the
difference of distances between the levels $C(S)$ and $C(O)$ and the least upper bound
$sup(C(S),C(O))$:
\[
p_1=-|dif(C(S),sup(C(S),C(O)))-dif(C(O),sup(C(S),C(O)))|\frac{m}{L}.
\]
If the level of confidence of the subject $C(S)$ and the level of secrecy of the object $C(O)$
are incomparable, access is not granted, so the value should be negative. In this case, because
the difference of "distances" between the levels of the lattice is defined, the absolute value
of the difference is calculated.

{\bf Example 2.} Consider the model example of such a system. Suppose that the mandatory security
policy is based on a nonlinear security lattice with eight levels
$L = {0, 1a, 1b, 1c, 2ab, 2c, 3, 4}$. In addition, $0 \leq 1a$, $0 \leq 1b$, $0 \leq 1c$
and levels $1a$, $1b$, $1c$ are incomparable, $1a \leq 2ab$, $1b \leq 2ab$, $1c \leq 2c$
and levels $2ab$, $2c$ are incomparable. Also $2ab \leq 3$, $2c \leq 3$, $3 \leq 4$.
For discretionary security policy, we assume that four types of access are defined:
$R = \{r, w, a, f\}$. Suppose that at some point in time a request for access $(s, o, r)$
is received, which has the corresponding cell $M[s, o] = \{r, w, a\}$ in the access matrix,
the level of secrecy of the object is $C(o) = 1c$, the level of confidence of the subject
is $C(s) = 2ab$. In this case, $sup(C(s), C(o)) = 3$, $diff(C(s), sup(C(s), C(o))) = 1$,
$diff(C(o), sup(C(s), C(o))) = 2$. Hence $p_1= -1 \times |1 - 2| = -1$ and $p_2 = 2$.
When security policies are equal, then $p = p_1 + p_2 = 1 > 0$, i.e. access is granted despite
the prohibition of the mandatory security policy. If we increase the priority of the mandatory
security policy 3 times, then:
\[
p=\frac{3}{4}\cdot(-1)+\frac{1}{4}\cdot2=-\frac{1}{4}<0.
\]
In this case, access is denied.

\section{Application of the analytic hierarchy process}

Two mandatory and two discretionary security policies often operate in the same system: 
one is related to confidentiality, the other is related to integrity. In this case, 
to calculate the clearance level it is more convenient to use the analytic hierarchy 
process with decision tree shown in Figure 1.

\begin{figure}[ht]
\centering
\includegraphics[width=0.5\textwidth]{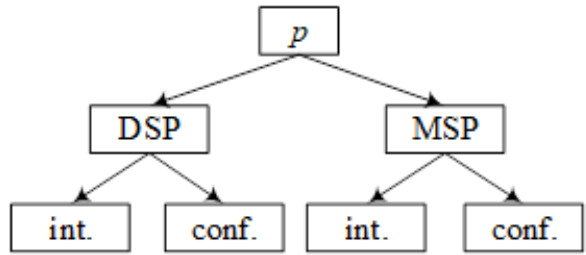}
\caption{The decision tree for calculating the clearance level $p$}
\label{fig1}
\end{figure}

It is necessary to fill three pairwise comparison matrices: one is for the level of criteria 
and two are for the level of alternatives. Suppose, as before, $r$ ($r > 0$) is the dominance
coefficient, showing how many times the decision taken by the mandatory security policy (MSP) 
is more important than the decision of the discretionary security policy (DSP). Preference 
for the confidentiality policy over the integrity policy is evaluated by two similar parameters: 
$r_1$ ($r_1 > 0$) for the discretionary model, $r_2$ ($r_2 > 0$) for the mandatory model. 
Then the pairwise comparison matrices are given in table:\\

\begin{tabular}{|l|l|l|}
\hline
$p$	&DSP	&MSP\\
\hline
DSP	&1	&$1/ r$\\
\hline
MSP	&$r$	&1\\
\hline
\end{tabular}
\begin{tabular}{|l|l|l|}
\hline
DSP	&int.	&conf.\\
\hline
int.&	1 &	$1/ r_1$\\
\hline
conf.&	$r_1$&	1\\
\hline
\end{tabular}
\begin{tabular}{|l|l|l|}
\hline
MSP	&int.	&conf.\\
\hline
int.	&1	&$1/ r_2$\\
\hline
conf. &	$r_2$&	1\\
\hline
\end{tabular}
\\

The ideal consistency of these matrices comes from the fact that for two-dimensional reciprocal
matrix $M$ there always is the condition: $\forall  i, j, k$ comes the equation 
$[M]_{ij} = [M]_{ik}\times [M]_{kj}$. In this case, the relative weight coefficients 
are determined by the normalized columns (for example, the first) of all three pairwise 
comparison matrices, and the formulas for calculating the relative priorities of the integrity policy and the confidentiality policy have the following forms:
\[
R^{int}=\frac{1}{1+r_1}\cdot\frac{1}{1+r}+\frac{1}{1+r_2}\cdot\frac{r}{1+r},\ \ 
R^{conf}=\frac{r_1}{1+r_1}\cdot\frac{1}{1+r}+\frac{r_2}{1+r_2}\cdot\frac{r}{1+r}=1-R^{int}.
\]
The final decision on granting the access can now be calculated according to the formulas:
\[
p=R^{int}\cdot p^{int}+R^{conf}\cdot p^{conf}.
\]
\[
p^{int}=\frac{1}{1+r}\cdot p^{int}_{DSP}+\frac{r}{1+r}\cdot p^{int}_{MSP},\ \ 
p^{conf}=\frac{1}{1+r}\cdot p^{conf}_{DSP}+\frac{r}{1+r}\cdot p^{conf}_{MSP}.
\]
where the superscript denotes the confidentiality policy or the integrity policy, 
and the subscript denotes the discretionary access control or the mandatory access control.
Analyzing the given formulas, we can draw the following conclusions:

1. Since $R^{int}$ and $R^{conf}$ belong to the range $(0, 1)$, then using the analytic hierarchy
process in case $p^{int}$ and $p^{conf}$ have the same sign will not change the decision of
granting the access.

2. If $r_1 \geq 1$ and $r_2 \geq 1$, then $R^{int} \leq R^{conf}$. If $r_1 < 1$ and $r_2 < 1$, 
then $R^{int} > R^{conf}$. In both cases, the analytic hierarchy process formulas can be replaced
by the formula
\[
p=\frac{1}{1+r'}\cdot p^{int}+\frac{r'}{1+r'}\cdot p^{conf},
\]
where $r'$ is the parameter characterizing how many times the decision taken 
by the confidentiality policy is more important than the integrity policy decision.

3. Application of the analytic hierarchy process gives the most interesting results 
in case $p^{conf}$ and $p^{int}$ have different signs and 
$((r_1 > 1) \wedge  (r_2 < 1)) \vee  ((r_1 < 1) \wedge  (r_2 > 1))$.

Access will be granted and the integrity policy will receive the priority.

Consider the other embodiment of a decision tree. Suppose the system still has two pairs 
of security policies. Now we consider the decision tree of the analytic hierarchy process 
shown in Figure 2.

\begin{figure}[ht]
\centering
\includegraphics[width=0.5\textwidth]{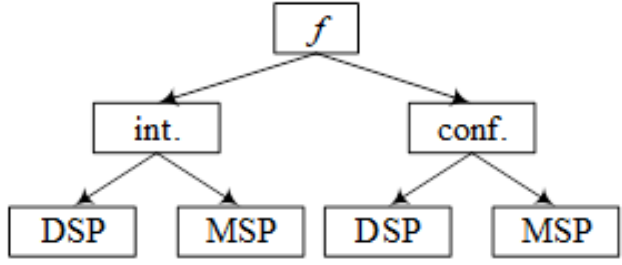}
\caption{The decision tree for calculating the clearance level $f$}
\label{fig2}
\end{figure}

Let the decision taken by the confidentiality policy $x$ times more important than 
the integrity policy decision. Preference for the mandatory access control model over 
the discretionary access control model is evaluated by two parameters: $x_1$ 
for the integrity policy, $x_2$ for the confidentiality policy. Then the pairwise 
comparison matrices are given in table:\\

\begin{tabular}{|l|l|l|}
\hline
F	&Int.	&Conf.\\
\hline
Int.	&1	&$1/ x$\\
\hline
Conf.&	$x$&	1\\
\hline
\end{tabular}
\begin{tabular}{|l|l|l|}
\hline
int.	&DSP	&MSP\\
\hline
DSP	&1	&$1/ x_1$\\
\hline
MSP	&$x_1$&	1\\
\hline
\end{tabular}
\begin{tabular}{|l|l|l|}
\hline
conf.	&DSP	&MSP\\
\hline
DSP	&1	&$1/ x_2$\\
\hline
MSP	&$x_2$	&1\\
\hline
\end{tabular}
\\

The formulas for calculating the relative priorities of the integrity policy and the confidentiality policy have the following forms:
\[
X_{DSP}=\frac{1}{1+x_1}\cdot\frac{1}{1+x}+\frac{1}{1+x_2}\cdot\frac{x}{1+x},\ \ 
X_{MSP}=\frac{x_1}{1+x_1}\cdot\frac{1}{1+x}+\frac{x_2}{1+x_2}\cdot\frac{x}{1+x}=1-X_{DSP}. 
\]
The final decision on granting the access can now be calculated according to the formulas:
\[
f=X_{DSP}\cdot f_{DSP}+X_{MSP}\cdot f_{MSP}.
\]
\[
f_{DSP}=\frac{1}{1+x}p_{DSP}^{int}+\frac{x}{1+x}p_{DSP}^{conf},\ \ 
f_{MSP}=\frac{1}{1+x}p_{MSP}^{int}+\frac{x}{1+x}p_{MSP}^{conf}.
\]
where the superscript denotes the confidentiality policy or the integrity policy, 
and the subscript denotes the discretionary access control or the mandatory access control. 
It is not difficult to prove the following statement.

{\bf Statement 1}. If $r = x_1 =x_2$ and $r_1 = r_2 = x$, then $p = f$.

Thus, in case of equality of priorities in the context of the selected access control model 
and in the context of confidentiality and integrity policies, both approaches 
to the construction of decision tree of the analytic hierarchy process lead to the 
same clearance level.

\section{Conclusion}
The proposed approach to the construction of a joint security policy has several advantages
compared to the traditional requirement of simultaneous access permission by all active security
policies. The presence of the weight coefficients allows the administrator to configure
the degree of influence of various security rules flexibly. Using two different types
of security policies make sense if they cover the different channels of information leakage.
Therefore, the choice of the weight coefficients in the decision algorithm must be based on
analysis of the probability of different attacks.

It should be noted that the need for a decision on the dominance of one security policy over
another arises only in case of conflict of permissions for the same access request. On the one
hand, in systems, which allow a consistent security administration, such conflicts do not arise.
On the other hand, if the two security policies never have a conflict, then one of the security
policies can be disabled without affecting the security of the system.

The proposed approach can be applied in the design of the additional information security
systems, as well as in software systems with their own security subsystem.



\begin{thebibliography}{9}

\bibitem{b1}
Bishop Ì. Applying the Take-Grant Protection Model. Technical Report. Dartmouth College Hanover, NH, USA, 1990.

\bibitem{b2}
Ribeiro C., Zuquete A., Ferreira P., Guedes P. SPL: An Access Control Language for Security Policies with Complex Constraints. In Proceedings of the Network and Distributed System Security Symposium, Sun Diego, CA, 2001.

\bibitem{b3}
Lunsford D. L., Collins M. R. The CRUD Security Matrix: A Technique for Documenting Access Rights. In Proceedings of the 7th Annual Security Conference, Las Vegas, NV, 2008.

\bibitem{b4}
Belim S., Bogachenko N., Ilushechkin E. An analysis of graphs that represent a role-based security policy hierarchy.Journal of Computer Security, vol. 23, no. 5, pp. 641-657, 2015

\bibitem{b6}
Kocaturk M. M., Gundema T. I. Fine-Grained Access Control System Combining MAC and RBAC Models for XML. Informatica, 2008, Vol. 19, Issue 4, pp. 517-534.

\end{thebibliography}
\end{document}